\documentstyle[epsf,aas2pp4]{article}
\begin{document}
\received{}
\revised{}
\accepted{}
\cpright{type}{year}
\journalid{VOL}{JOURNAL DATE}

\title{The orbital structure and potential of NGC 1399}

\author{R.P.~Saglia \altaffilmark{1}, Andi Kronawitter \altaffilmark{2},
Ortwin Gerhard \altaffilmark{2,3}, Ralf Bender \altaffilmark{1}}
\altaffiltext{1}{Institut f\"ur Astronomie und Astrophysik,
Scheinerstra\ss e 1, D-81679 Munich, Germany}
\altaffiltext{2}{Astronomisches Institut, Universit\"at Basel, Venusstrasse 7,
CH-4102 Binningen, Switzerland}
\altaffiltext{3}{Max-Planck-Institut f\"ur Astrophysik,
Karl-Schwarzschild-Str.~1, D-85740 Garching, Germany}

\begin{abstract}
Accurate and radially extended stellar kinematic data reaching $R=97''$ 
from the center are presented for the cD galaxy of Fornax, NGC 1399. 
The stellar rotation is small ($\le 30$ km/s); the stellar velocity 
dispersion remains constant at 250-270 km/s. The deviations from
Gaussian line of sight velocity distributions are 
small, at the percent level. 

We construct dynamical models of the galaxy, deprojecting its nearly
round (E0-E1) surface brightness distribution, and determining the
spherical distribution function that best fits (at the 4 percent
level) the kinematic data on a grid of parametrized potentials. We
find that the stellar orbital structure is moderately radial, with
$\beta=0.3\pm0.1$ for $R\le60''$, similar to results found for some
normal giant ellipticals. The gravitational potential is dominated by the
luminous component out to the last data point, with a mass-to-light
ratio $M/L_B=10~M_\odot/L_\odot$, although the presence of a central
black hole of $M
\approx 5 \times 10^8 M_\odot$ is compatible with the data in the
inner 5 arcsec. The infludence of the dark component
is marginally detected starting from $R\approx 60''$. Using the radial
velocities of the globular clusters and planetary nebulae of the
galaxy we constrain the potential more strongly, ruling out the
self-consistent case and finding that the best fit solution agrees
with X-ray determinations. The resulting total mass
and mass--to--light ratio are $M=1.2-2.5\times 10^{12} M_\odot$ and
$M/L_B=22-48~M_\odot/L_\odot$ inside $R=417''$ or 35 kpc for D=17.6
Mpc.
\end{abstract}


\section{Introduction}
\label{introduction}

Determining the gravitational potential of elliptical galaxies and the
amount of dark matter present is a notoriously difficult problem. This
stems from the absence of extended cold gas which could be used as
dynamical tracer, and the unknown intrinsic shape and orbital
structure (e.g., Saglia \markcite{r32} 1996 and references therein).
Progress has been made in recent years thanks to improved
observational data and dynamical modeling techniques.

On the one hand, we are now able to measure not only the stellar
velocity and velocity dispersion profiles, but also the shape
parameters of the line of sight velocity distributions with good
accuracy (van der Marel and Franx \markcite{r34} 1993; Bender, Saglia
and Gerhard \markcite{r4} 1994, hereafter BSG94; Carollo et al.\
\markcite{r35} 1995; Gerhard et al.\ \markcite{r10} 1998a, hereafter
G+98; Statler and Smecker-Hane \markcite{r56} 1999). Such data contain
information on the anisotropy and the gravitational potential (Gerhard
\markcite{r50} 1993, Merritt \markcite{r21} 1993).
On the other hand, algorithms based on Schwarzschild's (1979) orbit
superposition technique (Rix et al.\ \markcite{r25} 1997; Cretton et
al.\ \markcite{r44} 1999) or on the non-parametric reconstruction of
the underlying distribution function (\markcite{r10}~G+98; Matthias
and Gerhard \markcite{r45} 1999) allow a detailed modeling of the
data. 

Up to now, only a limited number of ellipticals have been analyzed
with these methods (Rix et al.\ \markcite{r25} 1997; \markcite{r10}
G+98; Matthias and Gerhard \markcite{r45} 1999).  Nevertheless, the
following pattern appears to emerge from these studies.  Out to 1-2
half-luminosity radii $R_e$ these objects are moderately radially
anisotropic, and they have a dark component which starts to dominate
the dynamics at about $R > 1-2 R_e$. This is consistent
with previous studies of extended velocity dispersion profiles of
ellipticals (Saglia, Bertin and Stiavelli \markcite{r38} 1992; Saglia
\markcite{r37} et al.\ 1993), of the radial velocities of globular
clusters (Mould et al.\ \markcite{r39} 1990; Grillmair et al.\
\markcite{r40} 1994; Cohen and Ryzhov \markcite{r41} 1997; Minniti et
al.\ \markcite{r20} 1998, Kissler-Patig et al.\ \markcite{r16} 1998)
and planetary nebulae (Arnaboldi et al. \markcite{r3} 1994, 1998; Hui et
al.\ \markcite{r54} 1995); of the X-ray emission around ellipticals
(Ikebe et al.\ \markcite{13} 1996, Jones et al.\ \markcite{r15} 1997)
and, more recently, the modeling of strong lensing systems (Romanowsky
and Kochaneck \markcite{r41} 1999) and the statistics of galaxy-galaxy
lensing (Geiger and Schneider \markcite{r54} 1998).

In this paper we apply the non-parametric distribution function
reconstruction method to new, accurate and extended data of the Fornax
cD galaxy NGC 1399, to explore whether the conclusions reached for
normal ellipticals are valid also for the class of the most massive
early-type galaxies. A preliminary description of the results can be
found in Kronawitter et al.\ \markcite{r54} (1999). In the following
we adopt a distance of 17.6 Mpc. NGC 1399 has been the subject of a
number of dynamical studies in the past. Bicknell et al.\
\markcite{r33} (1989) analyzed the major axis velocity dispersion
profile using the Jeans equations, concluding that a constant
mass-to-light ratio model with little anisotropy is able to reproduce
the data out to $\approx 85''$.  More recently, Graham et
al. \markcite{r42} (1998) reached the same conclusion using models
based on $R^{1/n}$ luminosity profiles. X-ray observations indicate
that the outer parts ($R>120''=10.2$ kpc) are dominated by a dark halo
which merges into the one of the Fornax cluster at around $R\approx
53$ kpc (Ikebe et al.  \markcite{13} 1996). However, the potential
inside $R\approx 10$ kpc is little constrained due to the presence of
a cooling flow (Rangarajan et al.\ \markcite{r24} 1995). The radial
velocities of globular clusters and planetary nebulae (see references
above) agree with these mass determinations.

The structure of the paper is as follows. We present the data in \S
\ref{observations}, where also the data reduction and the comparison
with the literature are described and discussed. We review our
dynamical modeling approach in \S \ref{modeling} and discuss some
modifications needed for the analysis of NGC 1399. Results are
presented in \S \ref{results}, where the constraints from the
absorption line data are combined with those from globular cluster and
planetary nebulae radial velocities and X-ray data.
Conclusions are drawn in \S \ref{conclusions}.

\section{Observations of NGC 1399}
\label{observations}

The observations were performed from the 6th to the 8th of December 1994,
using EMMI and the NTT, in remote observing mode from
Garching. The grating \# 6 (13.5 \AA /mm) was used in combination with the
ESO CCD Tek \#36 (2048$^2$ 24 $\mu$m pixels, 0.268 arcsec/pixel) and a
3 arcsec wide slit, giving a FWHM spectroscopic resolution of 3.5 \AA .
The central wavelength was 5150 \AA , giving 85 km/s $\sigma_{instr}$
resolution; the spectral range (4826-5474 \AA) covered the absorption
lines H$\beta$, Mg and Fe.  A 1.5 hour long spectrum was taken along
the major axis of the galaxy (PA=110$^0$); additional 9 hours of
integrations were taken parallel to the minor axis, shifting the
slit 42 arcsec from the center. Numerous template stars were also
observed, wiggled and trailed along the slit. 

The standard reduction steps (bias subtraction, flat fielding, cosmic
ray removal, logarithmic wavelength calibration, sky subtraction) were
performed under MIDAS. A sky subtraction better than 1 per cent was
achieved. The analysis of the data was carried out using the FCQ
method (Bender \markcite{r51} 1990) following BSG94 \markcite{r4} and
G+98 \markcite{r10}. The spectra were rebinned along the spatial
direction to obtain a nearly constant signal-to-noise ratio larger
than 50 per resolution element. The effects of the continuum fitting
and instrumental resolution were extensively tested by Monte Carlo
simulations. The residual systematic effects on the values of the
$h_3$ and $h_4$ profiles are less than 0.01, and less than 1.5\% in
$\sigma$. The errorbars 
(which reflect the random errors and do not take into account systematic 
effects such as template mismatching or the presence of dust and faint 
emission, see below) 
are very small, in the range 3-7 km/s for the
recessional velocities and velocity dispersion, and 0.006-0.02 for the
$h_3$ and $h_4$ coefficients. They are calibrated with simulations and
are determined to better than 20\%. They agree with the scatter of the
points except for the 5 pairs of points with $5<R<20$ arcsec. In this
region the spectra on the two sides of the galaxy are slightly
different, probably due to low level emission. The $H\alpha$ maps
(Goudfrooij \markcite{r12} et al.\ 1994, Singh \markcite{r28} et al.\
1995, Macchetto \markcite{r19 }et al.\ 1996) show that the emission is
patchy in the 10-20 arcsec region along the slit position.  Dust is
also present (Goudfrooij \markcite{r12} et al.\ 1994).  Bicknell
\markcite{r5} et al.\ (1989), who notice the same asymmetry, suggest
that "old shells" might also be partly responsible.  In the inner 1.5
arcsec, where $\sigma$ increases to 370 km/s, there is still some
template 
mismatching, which none of the 15 comparison stars is able to
eliminate.  The outermost datapoint has low signal and only the mean
velocity dispersion datapoints are reliable.

\begin{figure}
\begin{center}
\plotone{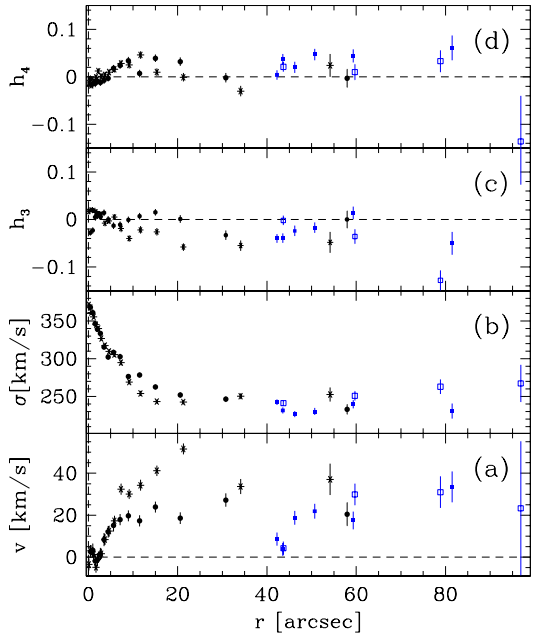}
\caption[kin1999]{\small The kinematics of NGC 1399. (a) The folded
mean velocity, (b) velocity dispersion, (c) $h_3$ and (d) $h_4$
profiles. Crosses and filled circles refer to the two sides of the
galaxy and the major axis spectrum.  Open and filled squares refer to
the two sides of the galaxy and the spectra taken parallel to the
minor axis and shifted 42 arcsec from the center.}
\label{kin1399}
\end{center}
\end{figure}

Fig. \ref{kin1399} shows the kinematical profiles including $h_3$ and
$h_4$, with the major axis data folded with respect to the center, and
the spectra parallel to the minor axis folded with respect to the
major axis. The rotation along the major axis is small, $\la 30$ km/s,
and is present also along the position parallel to the minor axis.
The velocities measured in the inner 3 arcsec might indicate the
presence of a kinematically decoupled core.  The velocity dispersion
increases to $\approx 370$ km/s in the inner 5 arcsec, and flattens to
$\approx 250$ km/s in the outer region.  The antisymmetric deviations
from a Gaussian profile as parametrized by the $h_3$ coefficients are
slightly negative where rotation is detected (see BSG94
\markcite{r4}). The symmetric deviations measured by the $h_4$
coefficients are smaller than 0.05 and mostly positive.

\begin{figure}
\begin{center}
\plotone{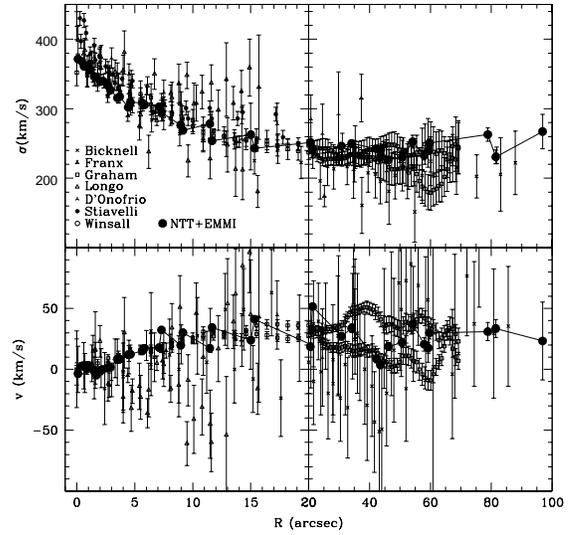}
\caption[compvel]{\small Comparison with the velocity dispersion (top)
and rotation velocities (bottom ) of
Bicknell \markcite{r5} et al.\ (1989, crosses), 
Franx \markcite{r9} et al.\ (1989, triangular diamonds),   
Graham et al.\ \markcite{r42} (1998, open squares), 
Longo \markcite{r18} et al.\ (1994, open triangles),
D'Onofrio \markcite{r7} et al.\ (1995, triangular crosses),
Stiavelli \markcite{r21} et al.\ (1993, stars),
Winsall and Freeman \markcite{r31} (1993, open circle).    
The line connects the data points determined here (full circles).
Note the change of scale at $R=20''$.
}
\label{compvel1399}
\end{center}
\end{figure}

The comparison with the rotational velocities along the major axis of
Bicknell \markcite{r5} et al.\ 1989), D'Onofrio \markcite{r7} et al.\
(1995), Franx \markcite{r9} et al.\ (1989), Graham \markcite{r11} et
al.\ (1998) and Longo \markcite{r18} et al.\ (1994) is shown in the
lower panel of Fig. \ref{compvel1399}.  There is overall good
agreement within the quoted errors. The systematic deviations observed
for the outermost datapoints of Longo et al.\ (1994) and D'Onofrio et
al.\ (1995) suggest that these datasets are probably affected by
residual sky subtraction errors. The comparison with the dispersion
profiles of Bicknell \markcite{r5} et al.\ (1989), D'Onofrio
\markcite{r7} et al.\ (1995), Franx \markcite{r9} et al.\ (1989),
Graham \markcite{r11} et al.\ (1998), Longo \markcite{r18} et al.\
(1994), Stiavelli \markcite{r21} et al. (1993), and Winsall and
Freeman \markcite{r31} (1993) is shown in the upper panel of
Fig. \ref{compvel1399}.  There is overall agreement within the
(usually large) errorbars given by the authors. Systematic differences
are observed in the inner 5 arcsec, where differences in seeing and
slit width may play an important role. In the outer parts the velocity
dispersions of Bicknell \markcite{r5} et al.\ (1989) are
systematically smaller than our data. However, our data agree well
with Graham et al.\ \markcite{r11} (1998) and with the point at
$R\approx 80$ arcsec measured by Winsall and Freeman \markcite{r31}
(1993).

\section{Dynamical modeling}
\label{modeling}

The algorithm used to analyse the kinematic data presented above
follows closely the technique discussed in detail by G+98. Therefore
here we limit ourselves to summarizing its main features and the
modifications we implemented to fit NGC 1399. In \S \ref{deprojection}
we focus on the modeling of the light distribution; in \S \ref{df} we
concentrate on the non-parametric reconstruction of the distribution
function.

\subsection{The deprojection of the light distribution}
\label{deprojection}

The surface-brightness (SB) data of NGC 1399 consist of three sets: in
the innermost part we use V-band HST-data from Lauer et
al. \markcite{r17} (1995), the medium range is covered by ground-based
B-band CCD-data from Bicknell \markcite{r5} et al.\ (1989), and the
outer part by photographic V-band data from Schombert \markcite{r26}
(1986).  As described in Bicknell \markcite{r5} et al.\ (1989) we used
the Schombert \markcite{r26} (1986) data shifted by 1.2 mag and from
57'' on outwards. The HST-data are also shifted by 0.8 mag and are
used out to 10''. With an effective radius of 42'' according to Faber
\markcite{r8} et al.\ (1989) the photometric data extend over 75
effective radii. 
(Note that Caon et al. \markcite{r100} 1994 find
$R_e\equiv\sqrt{a_e b_e}=127''$ by fitting
their extended photometry. The determination of the total magnitude and 
effective radius of NGC 1399 is highly 
uncertain, since its light profile is very shallow in the outer regions. The 
exact value of $R_e$ does not play a role in the following analysis).

The surface brightness profile deviates strongly from
a de Vaucouleurs or Jaffe profile, being much brighter in the outer
parts.  Thus contrary to NGC 6703, for which G+98 \markcite{r10} could
model the light distribution of NGC 6703 as a Jaffe \markcite{r14}
(1983) profile, we must here determine the three--dimensional luminous
density of NGC 1399 by deprojection.

NGC 1399 is nearly round (E0-E1). If we assume spherical symmetry the
deprojection of the SB profile $\Sigma(R)$ to a spatial luminosity
density $j^{lum}(r)$ is unique and is given by the standard Abel
equation
\begin{eqnarray}
\label{eqabel}
j^{lum}(r) = -\frac{1}{\pi} \int_{r}^{\infty} \frac{d \Sigma(R)}{d R} 
        \frac{d R}{\sqrt{R^2 - r^2}}.
\end{eqnarray}
To avoid the amplification of noise, we follow Wahba and Wendelberger
\markcite{r30} (1980), Scott \markcite{r27} (1990), and Gebhardt et
al.\ (1996), and first smooth the data by finding the function
$\Sigma(R)$ (expressed as thin plate splines) which maximizes
the logarithm $M$ of likelihood ${\cal L}$
\begin{eqnarray}
\label{eqlikesb}
M = \log {\cal L} = 
\sum_i \frac{\left(\log \Sigma_i - \log \Sigma(R_i) \right)^2}{\epsilon_i^2}
- \lambda P(\Sigma),
\end{eqnarray}
where $\Sigma_i$ denotes the observed SB at projected radius $R_i$ and
$\epsilon_i$ the measurement error.  The second term on the right hand
side is a penalty function $P$ (the second derivative of the SB
profile $-2.5 d^2 \log \Sigma/d\log R^2$) 
which penalizes oscillations in $\Sigma(R_i)$ induced by
noise.  The factor $\lambda$ controls the degree of smoothing. We
fixed it to $0.003$, which allows to reproduce smoothly the outer,
noisy regions of the galaxy without affecting the inner, high signal
to noise datapoints.
We also tried using the more objective Generalized
Cross Validation (GCV) method to determine the smoothing parameter,
but this did not give useful results for this inhomogeneous data set.

\begin{figure}
\begin{center}
\plotone{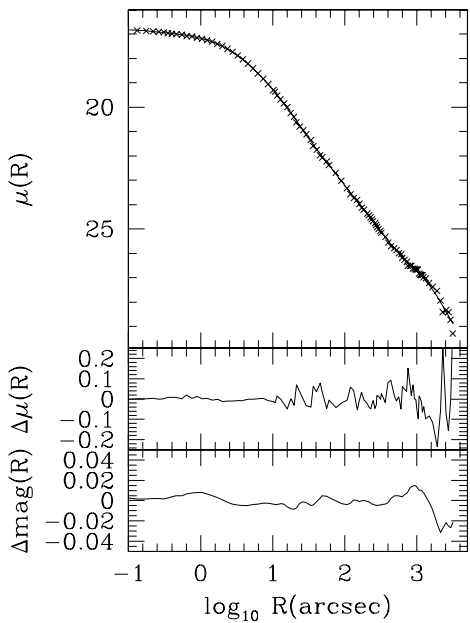}
\caption[dsb]{\small Top panel: the combined surface brightness
profile $\mu$ of NGC 1399 (crosses) and the projection of the derived
luminous density model (full line; see \S \ref{deprojection}). Middle
panel: the residuals $\Delta \mu$ in magnitudes per square arcsec
between the observed SB profile and the model. Lower panel: the
residuals $\Delta$mag in magnitudes between the integrated luminosity
of NGC 1399 and the model.}
\label{dsb}
\end{center}
\end{figure}

The smoothed SB-profile is then deprojected using eq.~(1).  The error
introduced by this inversion can be checked by generating data from a
known model and deprojecting.  For SB data drawn from a Jaffe-model
(Jaffe \markcite{r14} 1983), with similar quality as the NGC 1399
data, the deviations between the known and the deprojected luminosity
density profile are smaller than 
0.5\% rms.  
Figure \ref{dsb} shows the surface
brightness profile of NGC 1399 and that obtained by projecting the
luminosity density model computed from our deprojection method.  The
differences are smaller than 0.05 mag/arcsec$^2$ for most of the
radial range, and smaller than 0.02 mag in the integrated
luminosities.

The resulting density-profile is rather shallow, with some intrinsic
structure at large radii. This would be lost if we had fitted a
parametrized model to the surface brightness data; by contrast
the regression--inversion method described above does not smooth away
intrinsic features or force the profile to a fixed asymptotic decrease
in the outer parts, as is the case in many analytic models.  


\subsection{The distribution function}
\label{df}

To analyse the line of sight velocity profile data we proceed in two
steps (Merritt \markcite{r21} 1993, Rix \markcite{r25} et al.\ 1997,
G+98 \markcite{r10}): First, the gravitational potential is kept fixed
and the best-estimate distribution function (DF) is determined, in a
least--squares sense. For ideal data, the solution to this problem in
the spherical case would be unique (Dejonghe and Merritt \markcite{r6}
1992). Then, this process is repeated for different gravitational
potentials to determine those for which a statistically good fit to
the data can be achieved with positive DF. Because changing the
gravitational potential has a different effect on the line profiles
than changing the anisotropy of the stellar orbits (e.g., Gerhard et
al.\ 1998b), both can be constrained; i.e., only a limited range of
potentials is consistent with the data.

The potentials examined here are the sum of the potential of the
luminous matter, obtained by solving Poisson's equation for the
spherical luminosity density found in \S \ref{deprojection}, and that
of a dark halo taken to be
\begin{eqnarray}
\Phi^{DM} = \frac{1}{2} v_0^2 \ln \left(r^2 + r_0^2\right)
\label{eqdmpot}
\end{eqnarray}
where $v_0$ is the asymptotic circular velocity and $r_0$ the core
radius of the halo. We explored models with velocities $v_0$ from 0 
(the self-consistent model) to 780 km/s and $r_0$ from 0.5 
to 6.0 $R_e$. Models with these potentials minimize the
contribution of the dark halo near the center and hence maximize the
mass-to-light ratio of the luminous component. In this sense they are
similar to the maximum disk models for spiral galaxies; by analogy we
refer to them as {\sl maximum stellar mass models}.  A priori these
halo potentials might not be correct; in fact, from cosmological
simulations (Navarro, Frenk and White \markcite{r23} 1997) one expects
more centrally peaked halo distributions with asymptotic profiles at
large radii $\propto r^{-3}$. However, the mass profiles derived from
our approach match well the X-ray mass distribution (see Fig.~\ref{mr}
and \S \ref{mass}). In addition, the luminous density profile derived
in \S \ref{deprojection} is well approximated by a $r^{-2}$ power law
beyond 1''.

To recover the DF of a galaxy we construct a set of basis DFs in a
fixed potential. Each basis DF has a different anisotropy structure
and reproduces the luminous density profile determined as in 
\S\ref{deprojection}.  The total DF is then a weighted sum of 57 of
these basis functions:
\begin{eqnarray}
\label{eqdf}
f = \sum_k a_k f_k, 
\end{eqnarray}
with $\sum_k a_k = 1$ and the additional constraint that $\sum_k a_k
f_k \geq 0$. everywhere.
For the basis functions $f_k$ we used
sequences of tangentially anisotropic models as described in \S 4.1
and App.~A of G+98 \markcite{r10}, as well as the
isotropic model and a function with radially anisotropic
orbital structure.
The radial function is used for determining the optimal
velocity scales when computing the projected Gauss-Hermite moments for
each of the $f_k$. The coefficients $a_k$ are determined by fitting
the projected model kinematics to the $I$ datapoints such that
\begin{eqnarray}
\Delta^2=\sum_{k=1,I} \left\{ \chi^2_{\sigma,k}+\chi^2_{2,k} 
+\chi^2_{4,k}\right\}+ \lambda \sum_{i,j} \Lambda(f)_{i,j}
\label{eqchi}
\end{eqnarray}
is minimal. Here $\chi^2_\sigma$, $\chi^2_2$ and $\chi^2_4$ are as in
Eqs.~(8) and (9) of G+98 \markcite{r10} and measure the error-weighted
square differences between the velocity dispersions of the galaxy and
the model, and similarly for the second and fourth order Gauss-Hermite
moments related to the measured $h_4$ coefficients. Following G+98
\markcite{r10}, we write the penalty function $\Lambda$ as a
combination of the second derivatives of the composite DF, evaluated
on a grid of points ($E_i,x_j$) in the energy and circularity
integrals.
However, in the present case some care is needed in normalizing these
second derivatives, to avoid that a few data points dominate the
regularisation terms.  For, as expected from the correlation between
galaxy magnitude and the steepness of the central profile slope (Faber
et al.\ \markcite{r43} 1997), NGC 1399 has a rather shallow central
cusp. Flat density-profiles towards the center cause the distribution
function to become very steep (Dehnen \markcite{r42} 1993), and
similarly its second derivatives.
After testing different regularization functionals on simulated data
of kwown distribution 
functions, we found that the best results are  obtained by setting
\begin{eqnarray}
\Lambda(f)= \frac{1}{\left( f''_{\rm iso}\right)^2} 
		\left( \frac {\partial^2f} {\partial E^2} \right)^2 
            +\frac{2}{\left( f'_{\rm iso} \right)^2}
		\left( \frac {\partial^2f} {\partial E\partial x}
		\right)^2
	    +\frac{1}{\left( f_{\rm iso}\right)^2}
	  	\left( \frac {\partial^2 f} {\partial x^2} \right)^2,
\label{eqpenalty}
\end{eqnarray}
where $f_{\rm iso}$, $f'_{\rm iso}$ $f''_{\rm iso}$ are the isotropic 
distribution function and its first and second derivative with respect
to energy.

\begin{figure}
\centering
\plotone{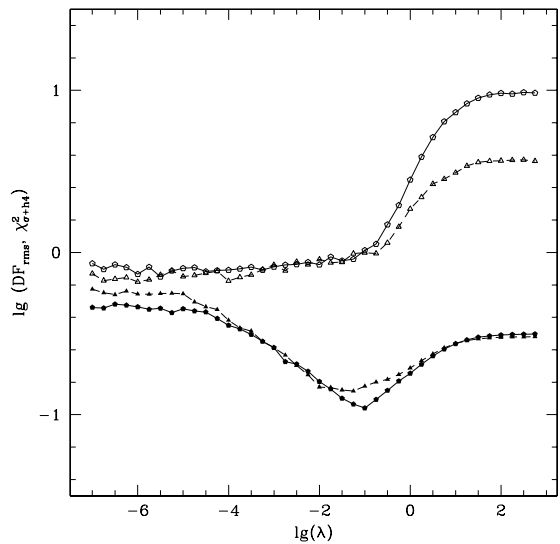}
\caption[residulambda]
 {\small Model results as a function of the regularization parameter
$\lambda$. The upper two curves show the total $\chi^2$ per $\sigma$ 
and $h_4$ data point of
regularized composite models with 57 basis functions. The lower 
curves show the rms deviation between the
recovered composite model distribution function and the true distribution 
function  from which the data
sets were generated. This rms deviation was evaluated on a grid in
energy and angular momentum corresponding to radii $< 3$ times
the radius of the outermost data point. Triangles refer to
the self-consistent model, pentagons to a model in a halo potential
 with $r_0=168$ arcsec and $v_0=427$ km/s.}
\label{residulambda}
\end{figure}

The value of $\lambda$ in Eq. \ref{eqchi} determines the degree of
smoothness of the composite DF. Low values of $\lambda$ result in a
perfect fit to the data at the price of unphysical small--scale
variations in the DF. Large values of $\lambda$ produce very smooth
DFs which might not give acceptable fits to the data. We estimate the
optimal value of $\lambda$ by considering the composite DF recovered
from Monte Carlo simulated data, which are derived from an underlying
distribution function similar to the one that gives the best fit to
the NGC 1399 data, with the assumption of Gaussian errors (see G+98 for
details).  Fig.~\ref{residulambda} shows the normalized total $\chi^2$
of the fit to the artificial data and the percentage rms variations of
the recovered DF, averaged over 25 realizations, as a function of
$\lambda$. A potential with only the luminous matter and one including
a halo with $r_0=168$ arcsec, $v_0=427$ km/s are used for
Fig.~\ref{residulambda}. In both cases values around
$\lambda\approx0.01$ give $\chi^2\approx 1$ and less than 15\% mean
rms variations of the DF.  In the analysis of NGC 1399 we thus adopt
$\lambda=0.02$.  Note that in G+98 we found that slightly different
values of $\lambda$ were needed to fit the self-consistent and the
halo potential cases. The difference here stems from the dominant role
of the luminous component in NGC 1399 (see \S \ref{mass}).

Finally, we constructed cumulative $\chi^2$ distributions for sets
of 100 Monte Carlo simulations of our data set, drawn from radial and
tangential models in self-consistent or halo potentials, respectively,
and analyzed using $\lambda=0.02$. We find that the 95\% confidence
level is reached at a value of $\chi^2\approx 1.3$ per data point in
all these cases. The same value results also when all errors are
increased by 25\%, with $\lambda=0.02$ still beeing the optimal
smoothing parameter. If we increase all errors by 40\%, the 95\%
confidence level is reached at a value of $\chi^2\approx 1.37$ per data
point. 

\section{Results}
\label{results}

\subsection{Stellar kinematics}
\label{starkin}

As discussed in \S \ref{observations}, the kinematic data show
differences between the two sides of the galaxy significantly larger
than the (small) statistical errors, probably due to low--level patchy
emission.  For radii $10<|R|<22$ arcsec, we therefore use in the
following the average values of $\sigma$ and $h_4$ over the two sides,
adopting as errors half of the side--to--side differences. In
addition, while minimizing the square differences between models and
datapoints over the whole radial range, we quote $\chi^2$ values
computed only from points at $|R|>5$ arcsec.  The velocity dispersion
of NGC 1399 steeply increases to 370 km/s in the central 5 arcsec,
which probably indicates the presence of a central massive object
(Stiavelli et al.\ \markcite{r21} 1993). We explore models with
central black holes at the end of this section.

\begin{figure}
\centering
\plotone{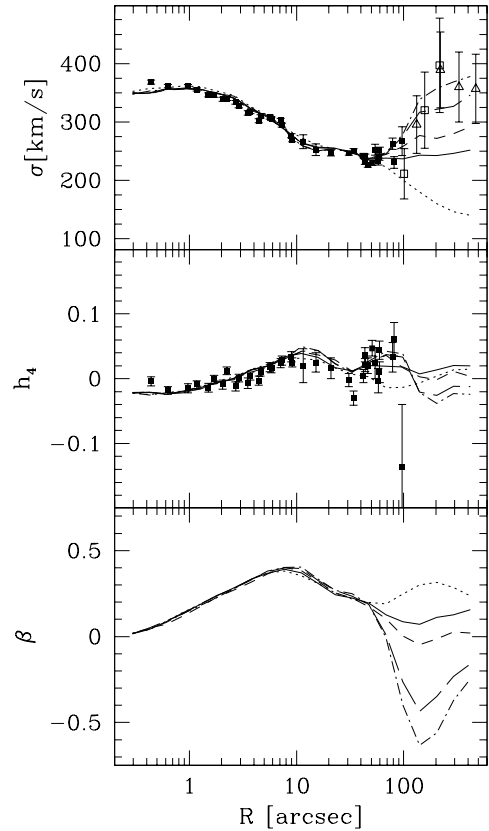}
\caption[Dynamical models for the kinematics of NGC 1399] {\small
Dynamical models for the kinematics of NGC 1399 in several luminous
plus dark matter potentials, compared to the measured projected
velocity dispersion (top panel) and velocity profile shape parameter
$h_4$ (middle panel). Full, dashed, long-dashed, dot-dashed lines: 
best-fit models with $V_{\rm circ}=383, 404,
425, 442$ km/s at $R=97''$ respectively. 
Dotted lines: self--consistent model with
luminous matter only, $V_{\rm circ}=364$ km/s.
Filled squares refer to the stellar
kinematics, open squares are the velocity dispersions derived for
globular clusters, reduced by the factor 1.08 (see \S \ref{globclus}).
The open triangles are the velocity dispersions derived from planetary
nebulae velocities. The bottom panel shows the models' intrinsic
anisotropy parameter $\beta(r)$. }
 \label{kinematicfits}
\end{figure}

The full line in Fig.~\ref{kinematicfits} shows a dynamical model
that best fits the kinematic data, obtained with a dark halo potential
with $r_0=210$ arcsec, $v_0=323$ km/s, corresponding to a total
circular velocity $V_{\rm circ}= 383$ km/s at the last data points,
$R=97''$.  Fits of similar quality are obtained
for a small range of model potentials; see the discussion in
G+98. Fig.~\ref{kinematicfits} shows three other models that span this
range in the present case. These have circular velocities at the last
data point of $V_{\rm circ}=404, 425, 442$ km/s, respectively. 
All these models represent the data very well, with  rms
differences in $\sigma$ of about 10 km/s, or 4.4 \%, and in $h_4$ of
about 0.018
(excluding the last datapoint). The
dotted line in the figure shows the best model without dark halo
($V_{\rm circ}=364$ km/s); this has similar rms differences in
$\sigma$ and $h_4$ when only datapoints within 60 arcsec are
considered, but substantially larger values, 19 km/s, or 8.7\% rms in
$\sigma$, and 0.027 rms in $h_4$, when the datapoints at $R>60''$ are
included. Therefore we conclude that the gravitational potential of
the galaxy in the region probed by the absorption line data is
dominated by the luminous component, but that the influence of the
dark component is marginally detected starting at $R\approx 60''$.

Despite the excellent fit,
the statistical interpretation of the resulting $\chi^2$ values is not
straightforward.  The top panel of Fig.~\ref{chisqvc} shows the values
of the normalized global $\chi^2_{\sigma+h_4}$ obtained for the model
fits, as a function of the circular velocity of the model at the last
datapoint, $R=97''$. Best-fit models are found in the range
$380<V_{\rm circ}<450$ km/s with values $\chi^2_{\sigma+h_4}\approx
1.9$, while the self-consistent model gives
$\chi^2_{\sigma+h_4}\approx 3.7$. The Monte Carlo simulations of our
data shown in Fig.~\ref{residulambda} demonstrate that the employed
set of basis functions is sufficient to produce $\chi^2\approx 1$, for
a Gaussian error distribution and when the underlying model
distribution function is similar to NGC 1399 and smooth.  In these
simulations the 95\% confidence limit corresponds to $\chi^2\approx
1.3$. The rather high $\chi^2$ values obtained for the real dataset
might partly be caused by an underestimation of the error bars.
Fitting to the data for NGC 1399 with all errorbars increased by 25\%
gives a normalized $\chi^2=1.3$ for our best--fitting model, just
marginally acceptable in a statistical sense (see above). Only when
all errorbars are increased by 40\% is the fitted model's $\chi^2$
reduced to 1.06. However, the (statistical) uncertainty of the Monte
Carlo--estimated error bars is not larger than 20\% (see \S
\ref{observations}), so most of the large $\chi^2$ must be due to
systematic effects.

Smaller, but still greater than unity, values of $\chi^2$ are obtained
if the two sides of the galaxies are fit separately ($\chi^2=1.19$ and
1.58 for the two sides, using the original errors). This indicates
that the systematic differences between the two sides discussed in \S
\ref{observations} are playing an important role. Also important are
sudden point--to--point variations in the measured data values which
are unlikely to be physical. We conclude that the residual differences
(at the two-three percent level) between model and data are of a
systematic nature, with the most obvious candidates being dust,
unresolved emission, local template mismatch, and slight oversmoothing
of the model. This also suggests that we might have reached the
intrinsic precision limit of stellar absorption line measurements.

To estimate confidence intervals for the potential parameters, we
therefore proceed as follows. We assume that due to systematic effects
the real errors on our kinematic data points are $40\%$ larger than
the statistical errors determined from Monte Carlo simulations, so
that the best-fitting model has a $\chi^2\simeq 1$ per data point (the
number of degrees of freedom is only slightly smaller than the number
of data points due to the enforced regularization). The $95\%$
confidence line is then at $\chi^2=1.37\times 1.4^2=2.68$ in units of
the original errors; see Figure \ref{chisqvc}, top panel. The
resulting $95\%$ confidence range for the circular velocity at
$R=97''$ is $V_{\rm circ}=420\pm 40$ km/s. This is consistent with the
sharp increase of the minimum values of $\chi^2$ seen for $V_{\rm
circ}<380$ km/s and $V_{\rm circ}>450$ km/s, which clearly indicates
that this procedure gives a fair estimate for the range of allowed
circular velocities for the combined luminous matter and halo
potential. In this sense the four halo models shown in
Fig.~\ref{kinematicfits} bracket the range of acceptable models.


The bottom panel of Fig.~\ref{kinematicfits} shows the anisotropy
profiles derived for the different potentials discussed above. As a
general trend, the profiles favour values of $\beta \approx 0.3$. The
maximum at 10'' and sharp drop in the inner 5 arcsec is uncertain (see
discussion below), while the tendency of the most massive models to
become tangentially anisotropic in the outer parts is also uncertain
until confirmed by data at still larger radii; see \S4.2 and the
discussion in G+98 \markcite{r10}. We conclude that NGC 1399 is
slightly radially anisotropic with $\beta=0.3\pm0.1$ out to $R=60$
arcsec or 5 kpc; and that the anisotropy is not yet well--constrained
at larger radii.  This is similar to what was found for NGC 6703 (G+98
\markcite{r10}) and NGC 2434 (Rix et al.\ \markcite{r25} 1997).

Finally, we briefly investigate the influence of the possible presence
of a massive black hole on the inferred anisotropy in the center,
without attempting a detailed analysis (this will be worthwhile only
for data with HST resolution).  We construct models with a central
black hole of mass around $M=5\times 10^8 M_\odot$, which is
approximately the mass expected for a galaxy of the luminosity of NGC
1399 (van der Marel \markcite{r49} 1999). We find that in these models
the fit to the velocity dispersion profile in the inner 5 arcsec
follows the steep increase of the measured $\sigma$ better, and the
anisotropy change is milder. In models with massive black holes the
inferred anisotropy can be substantially changed even at $10''$, and
is slightly reduced at $\sim 20''$.

\subsection{The Globular Cluster and Planetary Nebulae systems}
\label{globclus}

The mass distribution of NGC 1399 can be constrained further by
considering its globular cluster (GC) and planetary nebulae (PN)
systems.  Grillmair et al.\ \markcite{r40} (1994), Minniti et
al. \markcite{r20} (1998) and Kissler-Patig et al.\ \markcite{r16}
(1998) measured the radial velocities of 74 GCs, with typical errors
of 100 km/s. Forbes et al.\ \markcite{r46} (1998) studied the radial
distribution of the GCs and found that their surface density profile is
flatter than the galaxy starlight, following a power law of slope
$\gamma\approx -1.47$, 0.28 dex shallower than the galaxy. 
Thus at the same radius we expect the GCs to have a velocity
dispersion $\sim (1.75/1.47)^{1/2}\approx 1.08$ larger than the stars
of NGC 1399 (see, e.g., Binney and Tremaine \markcite{r47} 1987),
assuming a logarithmic potential and a nearly isotropic distribution
function for both components, which for the stars is approximately the
case. The discussion below does not change if we ignore this small
correction. 

Arnaboldi et al.\ \markcite{r3} (1994) measured the radial velocities
of 37 PNs, with typical errors of 50 km/s. The intrinsic density
distribution of the PNs is not as well established as for the GCs.  We
assume that in the outer parts of NGC 1399 the density distribution of
PNs follows that of the stars, with the same velocity dispersion. This
is consistent with the histogram shown in McMillian, Ciardullo and
Jacoby \markcite{r58} (1993), but the
number of PNs used is small and there is a possible selection bias
from the coupled effects of the metallicity dependence of the specific
density of PNs (Ciardullo and Jacoby \markcite{r55} 1992) and the
observed color gradients in elliptical galaxies (Richer, McCall,
Arimoto \markcite{r57} 1997). Fig.~\ref{kinematicfits} shows the
velocity dispersions (with their statistical errors) derived by
binning radially the GC and PN velocities.

We proceed to compute the likelihood of the two datasets on our grid
of gravitational potentials as follows. For each model we take the
projected velocity dispersion profile $\sigma(R)$, as determined in
\S\ref{starkin}, and approximate the probability for measuring a given
line--of--sight velocity at radius $R_i$ by a Gaussian of zero mean
and rms$=f \sigma(R_i)$, where $f=1.08$ for the comparison with the
GCs and $f=1$ for the PNs. This simplifying assumption is justified by
the low number of GCs and PNs velocities available and their rather
large errors. Monte Carlo simulations showed that deviations from
Gaussians can be detected only with samples of more than a few hundred
measured velocities; see also Merritt \markcite{r48} (1997).  After
convolution with the (Gaussian) error distribution specified by the
respective $\sigma_i$ of each data point, we write the likelihood of
the set of measured radial velocities $v_i$ as
\begin{equation}
{\cal L}_{GC}=\Pi_{i=1}^{74} 
\frac{1}{\sqrt{2\pi[f^2\sigma^2(R_i)+\sigma^2_i]}}
\exp\left( -\frac{v_i^2}{2[f^2\sigma^2(R_i)+\sigma^2_i]}\right),
\label{eqlikegc}
\end{equation}
and similarly for ${\cal L}_{PN}$. For comparison we also compute the
likelihoods ${\cal L}_{MC}$ of simulated sets of GC and PN radial
velocities, generated from our best-fit model of NGC 1399.

\begin{figure}
\centering
\plotone{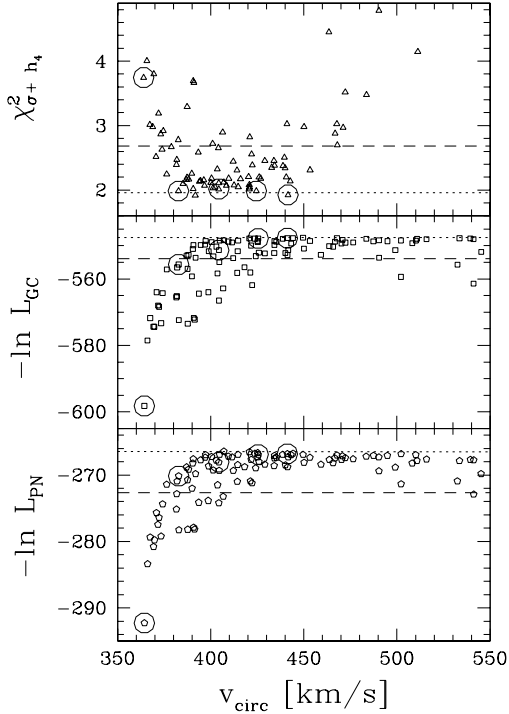}
 \caption[chisqvc]{\small Quality of the kinematic model fits for NGC
1399 as a function of the true circular velocity at $R=97''$. Top: the
normalized $\chi^2$ of the fit to the $\sigma-h_4$ datapoints. The
dotted line shows the mean $\chi^2$ (scaled to the real errors) of the
best-models obtained from Monte Carlo simulations, when the errors are
increased by 40\% (see text). The dashed lines show the related 95\%
level, indicating an allowed range of $V_{\rm circ}=420\pm 40$ km/s.
Middle: the likelihood $-\ln {\cal L}_{GC}$ of the Globular Cluster
dataset. Bottom: the likelihood $-\ln {\cal L}_{PN}$ of the Planetary
Nebulae dataset.  The models shown in Fig.~\ref{kinematicfits} are
circled. The dotted lines show the mean likelihood of the best-models
obtained from Monte Carlo simulations. The dashed lines show the 95\%
level.}
 \label{chisqvc}
\end{figure}

The middle and bottom panels of Fig.~\ref{chisqvc} show the
results. Here we plot as a function of the circular velocity at
$R=97''$ the likelihoods of the models. Additionally, we show two
lines.  The dotted line displays the mean value of the likelihood
$-\ln {\cal L}$ for 100 Monte Carlo realizations of $N=74$ GCs with
velocities drawn from the projected kinematics of one of our best
fitting models (-547.6), and the same quantity (-266.45) for $N=37$
PNs. The dashed line gives the 95\% confidence level below which only
$5\%$ of the Monte Carlo realizations fall, for the GC (-553.9) and
the PN (-272.5) samples. It is seen that the GCs and PNs favour the
high mass halo models of the range compatible with the absorption line
data, and are inconsistent with the self--consistent model.  The
models that are best for the GC sample have $-\ln {\cal L}_{GC}\approx
-547$; they therefore give a good representation of the available data
set.  The same is true for the PN set. For both the GCs and the PNs,
${\cal L}_{PN}$ decreases sharply for models with $V_{\rm circ}<380$ km/s, in
agreement with the results of \S \ref{starkin}.  The Monte Carlo
simulations quoted above rule out the self-consistent model to almost
certainty with either the GC or the PN sample. However, models with
more massive halos than allowed by the stellar kinematics are
compatible with the GC and PN data.


We have estimated the effect of a possible contamination of the GC and
PN samples by intracluster objects.  As suggested by Theuns and Warren
\markcite{r101} (1997) and Mendez et al.  \markcite{r102} (1997), some
of the GCs and PNs considered above could in fact be foreground or
background objects in the Fornax Cluster but not associated with NGC
1399.  To test the sensitivity of our results to such a contamination,
we repeat the analysis on a subset of GCs and PNs where the
10 or 20 percent
objects with the highest relative velocities have been eliminated.
This is an extreme case, since we might expect some of the contaminors
to have smaller velocities. We find that the results derived for the
GC set are rather robust, with the best-fit potential remaining within
the error range given above. However, when considering the PNs without
the 8 fastest objects dark halos with circular velocities as low as
$V_{circ}\approx 350$ km/s are allowed. Velocities $V_{circ}>550$ km/s
are also less favoured.

\subsection{The mass and M/L distribution}
\label{mass}

\begin{figure}
\centering
\plotone{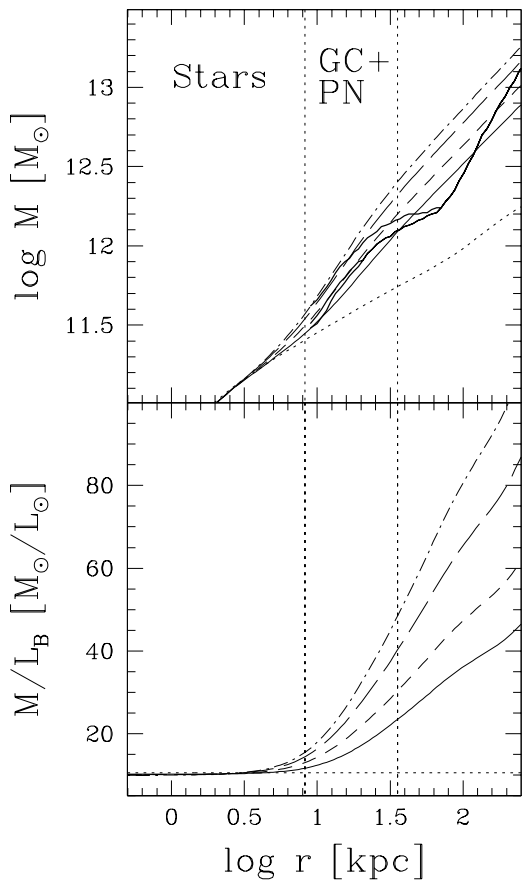}
 \caption[M(r) of dynamical models for NGC 1399]
    {\small Top: the total mass as a function of radius for the range
     of acceptable models for NGC 1399. Bottom: the mass to light ratio
    in the B band. Line styles as in Fig.~\ref{kinematicfits}. The full
    thick lines show the mass profiles derived from X-ray observations
    (see \S \ref{mass}).}
 \label{mr}
\end{figure}

Fig.~\ref{mr} shows the permitted range for the cumulative mass
distribution in NGC 1399, as derived from our combined fits to the
absorption line kinematics and the radial velocities of GCs and
PNs. The dark matter contribution is negligible for $R\le 60$ arcsec
or 5 kpc and becomes progressively dominant at larger $R$. It
comprises at most 1/3 of the total mass at the last stellar kinematic
data point, $R=97''$ or 8 kpc, and 1-3 times the luminous mass at the
radius of the most distant GC with a measured radial velocity,
$R\approx 417$ arcsec or 35 kpc. Inside this radius, the allowed mass
range matches the profiles derived from X-ray data by modeling the
cooling flow (Ikebe et al.\ \markcite{r13} 1996), giving
$M=1.2-2.5\times 10^{12} M_\odot$. At larger radii our simple models
based on logarithmic halo potentials fail to follow the knee inferred
from the X-ray data, which Ikebe et al. \markcite{r13} (1996)
interpret as the transition of the galaxy potential to the Fornax
cluster potential. Neverless these simple mass profiles are on the
right mass scale even at these large distances.

The integrated mass-to-light ratio in the B band is approximately
constant at $M/L_B\approx 10 M_\odot/L_\odot$ out to $R\approx 60''$,
increases to 12-15 at $R\approx 97''$ and reaches
$M/L_B=22-48~M_\odot/L_\odot$ at $R=417''$. Old ($t\ge 8$ Gyrs) and
metal rich ($Z\ge Z_\odot$) stellar populations (Worthey
\markcite{r55} 1994) can provide $M/L_B\approx 10$, matching at the
same time the values of the Mg$_2$ index and $(B-V)$ color observed in
the central regions of NGC 1399 (Mg$_2=0.334$, $(B-V)=0.99$, Faber et
al.\ \markcite{r8} 1989).

\section{Conclusions}
\label{conclusions}

We have presented new accurate and extended stellar kinematic data for
NGC 1399, the cD galaxy of the Fornax cluster. Using a non-parametric
algorithm, we have constructed smooth (at the 15 \% level) spherical
distribution functions which reproduce the observed velocity
dispersion and $h_4$ profiles at the four percent level.  The models
were tested for a grid of gravitational potentials, which combine a
luminous part obtained from the deprojected surface brightness
distribution with the assumption of constant M/L, and a
quasi--isothermal logarithmic dark halo potential. We have also
explored the influence of a central black hole of mass $5\times 10^8
M_\odot$, finding that the steep increase in $\sigma$ in the inner 5
arcsec can be better followed.  We have further constrained the mass
profile of the galaxy by taking into account the likelihoods of the
radial velocities measured for the galaxy's globular cluster and
planetary nebulae systems and by comparing the results with the
potential derived from X-ray measurements. Our main conclusions are
summarized as follows:

\begin{enumerate}

\item The stellar orbital structure of NGC 1399 is moderately radial
out to $R\approx 60''$, with $\beta=0.3\pm 0.1$, and is not yet
well--constrained at larger radii by the presently available data.

\item The luminous component of the galaxy dominates the dynamics out
to $R\approx 60''$, with $M/L_B=10 M_\odot/L_\odot$. Old and metal
rich stellar populations reproduce this high value, matching also the
measured values of the central $(B-V)$ color and Mg$_2$ index.

\item The influence of the dark halo is detected around
$R\approx 60-90''$ and becomes dominant in the region probed by the GC
and PN systems, where the self-consistent model is ruled out.

\item The circular velocity at the last measured stellar kinematic
point is $V_{\rm circ}=380-450$ km/s. The corresponding mass profile
range matches the mass curves derived from X-ray data with different
assumptions to model the galaxy's cooling flow. Inside $R=417''$ or 35
kpc we derive $M=1.2-2.5\times 10^{12} M_\odot$ and a cumulative
$M/L_B=22-48~M_\odot/L_\odot$.

\item The kinematics of the GCs and the PNs are consistent with those
of the stars within the large statistical errors. The GCs appear
slightly hotter,
even when a correction for the different density profile slope is
taken into account.
\end{enumerate}  

When compared to the normal giant ellipticals where a detailed stellar
dynamical analysis has been performed (NGC 1600, Matthias and Gerhard
\markcite{r45} 1999; NGC 2434, Rix et al.\ 1997 \markcite{r25}; NGC
6703, \markcite{r10} G+98), the cD galaxy NGC 1399 appears to have a
similar anisotropy profile inside the inner few kiloparsecs, with
$\beta\approx 0.3$. The luminous component of the galaxy dominates the
gravitational potential out to 8 kpc from the center. As judged from
the GC and PN data, the transition to the dark matter dominated regime
appears to be more rapid than in the case of NGC 6703. It is possible
that this difference reflects the special position of NGC 1399 at the
center of the Fornax cluster. The best way to derive more stringent
constraints on the nature of the transition region appears to be from
larger samples of PN velocities such as can be obtained with the VLT
and other large telescopes.

The present analysis based on spherical models is adequate to study
the very round inner regions of NGC 1399 considered here. For more
flattened systems the methods developed by Matthias and Gerhard
\markcite{r45} (1999) and Cretton et al.\ \markcite{r44} (1999) are
the future ways to follow.

\acknowledgments
We thank M.~Kissler-Patig for providing us with the globular
cluster velocities in electronic form, and the referee for
constructive comments. 
RPS and RB acknowledge the support by DFG grant SFB 375,
AK and OG by grant 20-50676.97 from the Schweizerischer Nationalfonds.
OG thanks the Max-Planck-Institut f\"ur Astrophysik for their
hospitality during a sabbatical visit.


\begin{references}
\reference{r3} 
Arnaboldi, M., Freeman, K.C., Hui, X., Capaccioli, M.,
Ford, H., 1994, The Messenger 76, 40
\reference{r59}
Arnaboldi, M., Freeman, K.C., Gerhard, O.E., Matthias, M., et al.,
1998, ApJ, 507, 759
\reference{r51}
Bender, R., 1990, A\&A, 229, 441
\reference{r4}
Bender, R., Saglia, R.P., Gerhard, O., 1994, MNRAS, 269, 785 (BSG94)
\reference{r5} 
Bicknell, G.V., Bruce, T.E.G., Carter, D., Killeen, N.E.B., 1989,
ApJ, 336, 639
\reference{r47}
Binney, J., Tremaine, S. 1987, Galactic Dynamics (Princeton: Princeton Univ.
Press)
\reference{r100}
Caon, N., Capaccioli, M., D'Onofrio, M., 1994, A\&AS, 106, 199
\reference{r35}
Carollo, C.M., de Zeeuw, P.T., van der Marel, R.P., Danziger, I.J., 
Qian, E.E., 1995, ApJ 441, L25
\reference{r54}
Ciardullo, R., Jacoby, G.H., 1992, ApJ, 388, 268
\reference{r41}
Cohen, J.G., Ryzhov, A., 1997, ApJ, 486, 230
\reference{r44}
Cretton, N., de Zeeuw P.T., van der Marel, R.P., Rix, H.-W., 1999, ApJ, 
submitted, astro-ph/9902034
\reference{r42}
Dehnen, W., 1993, MNRAS, 265, 250
\reference{r6}
Dejonghe, H., Merritt, D., 1992, ApJ, 391, 531 
\reference{r7}
D'Onofrio, M., Zaggia, S.R., Longo, G., Caon, N., Capaccioli, M., 
1995, A\&A, 296, 319
\reference{r8}
Faber, S., Wegner, G., Burstein, D., Davies, R.L., Dressler, A., 
Lynden-Bell, D., Terlevich, R., 1989, ApJS, 69, 763 
\reference{r43}
Faber, S.M., Tremaine, S., Ajhar, E., Byun, Y.-I., Dressler, A.,
Gebhardt, K., Grillmair, C., Kormendy, J., Lauer, T., Richstone, D. 
1997, AJ, 114, 1771
\reference{r46}
Forbes, D.A., Grillmair, C.J., Willinger, G.M., Elson, R.A.W., Brodie, J.P., 
1998, MNRAS, 293, 325
\reference{r9} 
Franx, M., Illingworth, G., Heckman, T., 1989, ApJ, 344, 613
\reference{r22} 
Gebhardt, K., et al., 1996, AJ, 112, 105
\reference{r54}
Geiger, B., Schneider, P., 1998, MNRAS, 302, 118
\reference{r50}
Gerhard, O.E., 1993, MNRAS, 265, 213
\reference{r10}
Gerhard, O.E., Jeske, G., Saglia, R.P., Bender, R., 1998a, 
MNRAS, 295, 197 (G+98)
\reference{r52}
Gerhard, O.E., Jeske, G., Saglia, R.P., Bender, R., 1998b, in
Galactic Halos, ASP Conf.~Ser.~Vol.~136, ed.\ Zaritsky, D., ASP,
San Francisco, 248
\reference{r12}
Goudfrooij, P., Hansen, L., J\o rgensen, H.E., N\o rgaard-Nielsen, H.U.,
1994, A\&AS, 105, 341
\reference{r11}
Graham, A., Colless, M.M., Busarello, G., Zaggia, S., Longo, G., 1998,
A\&AS, 133, 325
\reference{r42}
Graham, A., Colless, M., Busarello, G., 1998, ASP 136, 257
\reference{r40}
Grillmair, C.J., Freeman, K.C., Bicknell, G.V., Carter, D., Couch, W.J., 
Sommer-Larsen, J., Taylor, K., 1994, ApJ, 422, L9
\reference{r54}
Hui, X., Ford, H.C., Freeman, K., Dopita, M.A., 1995, ApJ, 449, 592
\reference{r13}
Ikebe, Y., Ezawa, H., Fukazawa, Y., Hirayama, M., Ishisaki, Y., Kikuchi, K., 
Kubo, H., Makishima, K., Matsushita, K., Ohashi, T., Takahashi, T.,
Tamura, T., 1996, Nature, 379, 427
\reference{r14}
Jaffe, W., 1983, MNRAS, 202, 995 
\reference{r15}
Jones, C., Stern, C., Forman, W., Breen, J., David, L., Tucker, W., Franx, M., 
1997, 482, 143
\reference{r16}
Kissler-Patig, M., Brodie, J.P., Schroder, L.L., Forbes, D. A., 
Grillmair, C.J., Huchra, J.P., 1998, ApJ, 115, 105
\reference{r53}
Kronawitter, A., Gerhard, O., Saglia, R.P., Bender, R., 1999, ``Dynamical
analysis of elliptical galaxy halos'', Proc. of the Conference
``Galaxy dynamics'', August 1998, Rutgers, ASP, in press
\reference{r17}
Lauer, T., Ajhar, E., Byun, Y., Dressler, A., Faber, S.,
Grillmair, C., Kormendy, J., Richstone, D., Tremaine, S., 1995, ApJ, 110, 2622 
\reference{r18}
Longo, G., Zaggia, S.R., Busarello, G., Richter, G., 1994, A\&AS, 105, 433 
\reference{r19}
Macchetto, F., Pastoriza, M., Caon, N., Sparks, W.B., Giavalisco, M.,
Bender, R., Capaccioli, M., 1996, A\&AS, 120, 463
\reference{r45}
Matthias, M.,  Gerhard, O.~E. 1999, MNRAS, in press (astro-ph/9901036)
\reference{r58}
McMillian, R., Ciradullo, R., Jacoby, G.H., 1993, ApJ, 416, 62
\reference{r102}
Mendez, R.H., Guerrero, M.A., Freeman, K.C., Arnaboldi, M., Kudritzki, R.P., 
Hopp, U., Capaccioli, M., Ford, H., 1997, ApJ, 491, L23
\reference{r21}
Merritt, D., 1993, ApJ, 413, 79 
\reference{r48} 
Merritt, D., 1997, AJ, 114, 228
\reference{r20}
Minniti, D., Kissler-Patig, M., Goudfrooij, P., Meylan, G., 1998, ApJ, 115, 121
\reference{r39}
Mould, J., Oke, J.B., de Zeeuw, P.T., Nemec, J.M., 1990, AJ, 99, 1823
\reference{r23}
Navarro, J.F., Frenk, C.S., White, S.D.M., 1997, ApJ, 490, 493 
\reference{r24}
Rangarajan, F.V.N., Fabian, A.C., Forman, W.R., Jones, C., 1995, 
MNRAS, 272, 665
\reference{r57}
Richer, M.G., McCall, M.L, Arimoto, N., 1997, A\&AS, 122, 215
\reference{r25}
Rix, H.-W., de Zeeuw, P.T., Cretton, N., van der Marel, R., Carollo, M.C., 
1997, ApJ, 488, 702 
\reference{r41}
Romanowsky, A.J., Kochaneck, C.S., 1999, ApJ, 516, 18
\reference{r38}
Saglia, R.P., Bertin, G., Stiavelli, M., 1992, ApJ 384, 433
\reference{r37}
Saglia, R.P., Bertin, G., Bertola, F., Danziger, I.J., Dejonghe, H., 
Sadler, E.M., Stiavelli, M., de Zeeuw, T., Zeilinger, W.W., 1993, ApJ, 403, 567
\reference{r32}
Saglia, R.P., 1996, Proc. of the IAU 171 Symposium 
``New light on Galaxy Evolution'', Heidelberg, Germany,
R. Bender, R.L. Davies Eds, p. 157
\reference{r26} 
Schombert, J.M., 1986, ApJS, 60, 603
\reference{r101}
Schwarzschild, M., 1979, ApJ, 232, 236
\reference{r27}
Scott, D.W., 1990, Multivariate Density Function, Wiley \& Sons, New York
\reference{r28}
Singh K.P., Bhat P.N., Prabbu T.P., Kembhavi A.K., 1995, A\&A, 302, 658
\reference{r56}
Statler, T.S., Smecker-Hane, T., 1999, AJ, 117, 839 
\reference{r21} 
Stiavelli, M., M\o ller, P., Zeilinger, W.W., 1993, A\&A, 277, 421
\reference{r101}
Theuns, T., Warren, S.J., 1997, MNRAS, 284, L11
\reference{r34}
van der Marel, R., Franx, M., 1993, ApJ, 407, 525
\reference{r49}
van der Marel, R., 1999, Proceedings of the Conference ``Galaxy Dynamics'', 
D. Merritt, M. Valluri, J. Sellwood, Eds., August 1998, Rutgers,
ASP in press, astro-ph/9811025
\reference{r30} 
Wahba, G., Wendelberger, J., 1980, Monthly Weather Review, 108, 1122
\reference{r31}
Winsall, M.L, Freeman, K.C., 1993, A\&A, 268, 443
\reference{r55}
Worthey, G. 1994, ApJSS, 95, 107
%
\end{references}
\end{document}